\documentclass[fleqn,10pt]{wlscirep}
\usepackage[utf8]{inputenc}
\usepackage[T1]{fontenc}
\usepackage{bm}
\usepackage{upgreek}
\usepackage[displaymath,mathlines]{lineno}
\DeclareMathAlphabet{\mathcal}{OMS}{cmsy}{m}{n}

\title{Intrinsic energy spread and bunch length growth in plasma-based accelerators due to betatron motion}

\author[1,2,*]{A. Ferran Pousa}
\author[1]{A. Martinez de la Ossa}
\author[1]{R. W. Assmann}
\affil[1]{Deutsches Elektronen-Synchrotron DESY, Hamburg, 22607, 
Germany }
\affil[2]{Institut f\"ur Experimentalphysik, Universit\"at Hamburg, Hamburg, 22761, Germany}
\affil[*]{angel.ferran.pousa@desy.de}

\begin{abstract}
Plasma-based accelerators (PBAs), having demonstrated the production of GeV electron beams in only centimetre scales, offer a path towards a new generation of highly compact and cost-effective particle accelerators. However, achieving the required beam quality, particularly on the energy spread for applications such as free-electron lasers, remains a challenge. Here we investigate fundamental sources of energy spread and bunch length in PBAs which arise from the betatron motion of beam electrons. We present an analytical theory, validated against particle-in-cell simulations, which accurately describes these phenomena. Significant impact on the beam quality is predicted for certain configurations, explaining previously observed limitations on the achievable bunch length and energy spread. Guidelines for mitigating these contributions towards high-quality beams are deduced.
\end{abstract}
\begin{document}

\flushbottom
\maketitle
%
%
\thispagestyle{empty}


\section*{Introduction}

In plasma-based accelerators (PBAs), an intense laser pulse \cite{PhysRevLett.43.267} or high-energy charged particle beam \cite{PhysRevLett.54.693} drives a plasma wake sustaining accelerating fields orders of magnitude higher than those achievable with conventional radiofrequency technology \cite{faure2006controlled}, offering a path towards highly compact and cost-effective accelerators. Having reached GeV energies in only centimeter scales \cite{leemans2006gev, wang2013quasi, leemans2014multi,mirzaie2015demonstration, PhysRevLett.122.084801}, femtosecond-long electron bunches with kiloampere current \cite{lundh2011few,couperus2017demonstration} and micron-level emittance \cite{fritzler2004emittance,brunetti2010low}, PBAs are becoming increasingly attractive for applications such as free-electron lasers (FELs) \cite{madey1971stimulated} or future linear colliders \cite{leemans2009laser}, which would strongly benefit from reduced size and cost. In particular, there is also a special interest in the production of sub-femtosecond bunches \cite{li2013dense, WEIKUM201633, PhysRevLett.119.044801} to generate short X-ray pulses for ultrafast science \cite{zewail2003atomic}.

However, despite these major advances, achieving an energy spread below the percent level has remained an issue. This is a problem not only for applications such as FELs, which require a relative energy spread $\lesssim 10^{-3}$ \cite{corde2013femtosecond}, but also for the beam transport after the plasma, which could cause a dramatic emittance growth \cite{migliorati2013} and further reduce the beam quality, limiting any kind of multi-stage acceleration. Although the impact on the emittance can be mitigated by tailored plasma-to-vacuum transitions \cite{dornmair2015} or the use of active plasma lenses instead of conventional quadrupoles \cite{vanTilborg2015}, improving the beam parameters in the plasma is essential for demanding applications. Major efforts have therefore been dedicated towards this objective, including the development of controlled injection techniques \cite{esarey1997electron, oz2007ionization, schmid2010density, PhysRevLett.111.245003, PhysRevLett.108.035001} as well as concepts for mitigating the correlated energy spread arising from the steep slope of the accelerating fields. These include beam loading the wake \cite{katsouleas1987beam,doi:10.1063/1.1889444, tzoufras2008beam}, either by the accelerated bunch itself \cite{doi:10.1063/1.4929921} or the injection of a secondary  one\cite{manahan2017single}, the use of modulated \cite{brinkmann2017chirp} or tailored \cite{PhysRevLett.121.074802} plasma profiles, the insertion of a magnetic chicane in a multistage PBA \cite{PhysRevLett.123.054801}, or taking advantage of the beam-induced wakefields \cite{BANE2012106,PhysRevLett.112.114801, PhysRevLett.122.034801, PhysRevLett.122.114801, PhysRevLett.122.204804}.

Here we describe additional sources of energy spread in PBAs which should also be mitigated, as well as limitations on the minimum achievable bunch length. In particular, we show that the transverse oscillations of beam electrons (known as betatron motion), induced by the strong focusing fields in the plasma wake, can be a significant source of uncorrelated (or slice) energy spread as well as increased bunch length due to path length differences between particles with different oscillation amplitude. Although these effects were already known \cite{reitsma2004coupling,xu2014phase}, no model exists to date for their impact on the beam parameters. We present here a novel analytical theory, validated against numerical simulations with the particle-in-cell (PIC) codes OSIRIS \cite{fonseca2002osiris} and HiPACE \cite{Mehrling_2014}, which accurately describes these phenomena in the assumption of relativistic electrons in a non-evolving wake. This model allows us to understand previously observed limitations, such as a finite energy spread even when the initial bunch length approaches zero \cite{grebenyuk2014laser}, and define guidelines for minimizing the impact of these phenomena for high-quality electron beams.

\section*{Results}

\subsection*{Development of a single-particle model}
As a first step, a single-particle model capable of describing the coupling of the transverse oscillations into the longitudinal motion is needed. Then, from this model, statistical averages over a particle beam can be analytically obtained in order to determine the impact of this coupling on parameters such as the energy spread and bunch length.

For a relativistic particle moving close to the speed of light $c$, the longitudinal component of its velocity $v_z = (v^2-v_x^2-v_y^2)^{1/2}$, where $v=|\bm{v}|$ is the magnitude of the particle velocity $\bm{v}$, is limited by $c$ and can be decreased due to motion in the transverse planes, i.e, if $v_x,v_y \neq 0$. Thus, in the speed-of-light frame $\xi = z - ct$, where $z$ is the longitudinal coordinate in the laboratory frame and $t$ is the time, a particle performing betatron oscillations will experience a slippage towards the back which, at a time $t$, will be given by $\Delta\xi(t) = \xi(t)-\xi_0 = \int_0^t v_\xi(t) dt$, where $\xi_0$ is its initial position and $v_\xi = v_z - c$ is its longitudinal velocity in the speed of light frame.

Determining the particle slippage thus requires obtaining an expression for $v_\xi(t)$, which, in turn, depends on the evolution of the transverse components $v_x$ and $v_y$. In order to find these expressions, the equations of motion of a relativistic electron in a plasma wakefield have to be solved. These are given by $\dot{\bm{p}} = -e \bm{W}$, where $\bm{p}= m \gamma \bm{v}$ is the particle momentum, $m$ is the electron mass, $\gamma=1/\sqrt{1-(\bm{v}/c)^2}$ is the relativistic Lorentz factor of the particle, $e$ is the elementary charge and $\bm{W}= \left(E_x-cB_y, E_y + cB_x, E_z\right)$ is the wakefield. In this expression, $E_i$ and $B_i$ (for $i=x,y,z$) are the different components of the electric and magnetic fields. Analytical expressions of the wakefields can be found for the linear regime \cite{gorbunov1987excitation, sprangle1988laser}, and several models have been developed for non-linear wakes in the blowout regime \cite{lotov2004blowout, lu2006nonlinear}. In what follows, the wakefield gradients around the particle, $K_x = \partial_xW_x$, $K_y = \partial_yW_y$ and $E_z'= \partial_zW_z$, will be assumed constant and $\partial_iW_j=0$ for $i\neq j$. This is especially well-suited for the blowout regime, where the driver expels all plasma electrons and leaves behind an ion cavity with uniform focusing fields $K_x=K_y= m\omega_p^2/2e$ and an approximately constant $E_z'$ in a wide range of the accelerating phase. The plasma frequency is defined as $\omega_p = \sqrt{n_pe^2/m \epsilon_0}$, where $\epsilon_0$ is the vacuum permittivity and $n_p$ is the unperturbed plasma density. For linear wakes, although these gradients are not uniform, this model can be applied for regions sufficiently close to the propagation axis, where $K_x$ and $K_y$ can be regarded as constant, and if $\Delta\xi \ll c/\omega_p$, so that the longitudinal change in $K_x$, $K_y$ and $E_z'$ can be neglected.

Under these conditions, the equations of motion for a beam electron in a non-evolving wake with a phase velocity $v_w$ given by the propagation velocity of the driver, can be written as
\begin{align}
\dot{\gamma}(t) &= \mathcal{E}(t), \label{eq:lon_eq} \\
\ddot{p}_x(t) + \omega_{x}(t)^2\,p_x(t) & =0, \label{eq:transv_eq}
\end{align}
where $\mathcal{E}(t) = -eE_{z}(t)/m c$, $\omega_x(t) = \sqrt{\mathcal{K}_x/\gamma(t)}$ is the betatron frequency and $\mathcal{K}_x = eK_x/m$. Since $\mathcal{E}'=-eE_z'/mc$ is assumed constant, then $\mathcal{E}(t) = \mathcal{E}_0 + \mathcal{E}'\left[\Delta\xi(t) + (c-v_w)t\right]$, with $\mathcal{E}_0=\mathcal{E}(0)$. Using this expression in Eq. (\ref{eq:lon_eq}) yields $\gamma(t) = \gamma^{(0)}(t) + \gamma^{(1)}(t)$, where $\gamma^{(0)}(t) = \gamma_0 + \mathcal{E}_0 t$, with $\gamma_0 = \gamma(0)$, is the energy evolution assuming constant acceleration and $\gamma^{(1)}(t) = \int_{0}^{t}\mathcal{E}'\left[\Delta \xi(t') + (c-v_w)t'\right]dt'$ accounts for the contribution of slippage and dephasing with respect to the wake \cite{schroeder2011nonlinear}. If $\gamma^{(1)}/\gamma^{(0)} \ll 1$,  then $\gamma \simeq \gamma^{(0)}$ can be assumed in order to solve Eq. (\ref{eq:transv_eq}), for which analytical solutions can be found if $\omega_x(t)$ is a slowly varying function \cite{kostyukov2004phenomenological}, i.e, if $ \dot{\omega}_{x}/\omega_{x}^2 = \mathcal{E}_0/2\sqrt{\gamma \mathcal{K}_x} \ll 1$. This yields the following expressions:
\begin{align}
x(t) &\simeq A_{x,0} \Gamma(t)^{-1/4}\cos{\left(\phi_x(t) + \phi_{x,0}\right)} , \label{eq:x} \\
v_x(t) &\simeq -A_{x,0} \omega_{x,0} \Gamma(t)^{-3/4}\sin{\left(\phi_x(t) + \phi_{x,0}\right)} , \label{eq:vx}
\end{align}
where $\Gamma(t) = \gamma^{(0)}(t)/\gamma_{0}=1+\mathcal{E}_0t/\gamma_{0}$ is the relative energy evolution, $A_{x,0} = \sqrt{x_0^2 + \left(v_{x,0}/\omega_{x,0}\right)^2}$ is the initial oscillation amplitude in the $x$ plane and $\phi_{x,0} = -\arctan\left(v_{x,0}/x_0 \omega_{x,0}\right)$ is the initial phase. The values $x_0$, $v_{x,0}$ and 
$\omega_{x,0}$ correspond to the initial transverse position, velocity and betatron frequency. The phase advance $\phi_x(t) = \int_0^t \omega_x(t')\, \mathrm{d}t'$ is given by
\begin{equation}\label{eq:mu}
\phi_x(t) \simeq 2\frac{ 
	\sqrt{\mathcal{K}_x \gamma_{0}}}{\mathcal{E}_0}\left(\Gamma(t)^{1/2} - 1\right) .
\end{equation} 
Analogous expressions can be found for the $y$ plane and are not included here for simplicity.

The solutions to the transverse electron motion can now be used to obtain the longitudinal slippage, $\Delta\xi(t) = \int_0^t (v_z(t') - c) \, \mathrm{d}t'$,
by considering $v_z = (\bm{v}^2-v_x^2-v_y^2)^{1/2} \simeq |\bm{v}|\,-(v_x^2+v_y^2)/2|\bm{v}|$ and $|\bm{v}|-c \simeq -c/2\gamma^2$
for a relativistic electron.
Since the time scale of the oscillation amplitude damping, $\Gamma^{-3/4}(t)$, in Eq.~(\ref{eq:vx}) is much longer than the betatron period, the time-average of this expression can be used for the electron transverse velocity. This results in the expression
\begin{equation}\label{eq:finalXi}
\Delta\xi(t) \simeq \frac{c}{2\mathcal{E}_0\gamma_0} \left(\Gamma(t)^{-1}-1\right) + \frac{A_0^2 \mathcal{K} }{2c\mathcal{E}_0}\big(\Gamma(t)^{-1/2}-1\big),
\end{equation}
where $A_0^2 = A_{x,0}^2 + A_{y,0}^2$ and the focusing gradient is assumed to 
be the same in both planes, i.e $\mathcal{K}_x=\mathcal{K}_y=\mathcal{K}$. The 
two terms in Eq. (\ref{eq:finalXi}) account, respectively, for the slippage due to $|\bm{v}| < c$ and the slippage caused by betatron motion. Most of $\Delta\xi$ occurs initially (at lower energies) and reaches a limit $\Delta\xi_{\mathrm{max}}$ as $\Gamma \to \infty$ given by
\begin{equation}\label{eq:xiLim}
\Delta\xi_{\mathrm{max}} = 
-\frac{1}{2\mathcal{E}_0}\left(\frac{c}{\gamma_0} 
+ \frac{A_0^2 \mathcal{K}}{c} \right) .
\end{equation}

Using Eq. (\ref{eq:finalXi}) it is now possible to calculate $\gamma^{(1)}$ and finally obtain  $\gamma(t) = \gamma^{(0)}(t) + \gamma^{(1)}(t)$, which takes into account the influence of the particle slippage on its energy. This yields
\begin{equation}\label{eq:finalGamma2}
\begin{split}
\gamma(t) \simeq \gamma_0 + \mathcal{E}_0t + \mathcal{E}'\Delta\xi_{\mathrm{max}}t +
\frac{\mathcal{E}'}{2}\left(c-v_w\right)t^2  
+ \frac{c\mathcal{E}'}{2\mathcal{E}_0^2}\ln{\Gamma(t)} 
+ \frac{\mathcal{E}'A_0^2\mathcal{K}\gamma_{0}}{c\mathcal{E}_0^2}\left(\Gamma(t)^{1/2} - 1\right) ,
\end{split}
\end{equation}
where the third term is a correction to the linear energy gain, which is modified due to slippage, and the fourth one accounts for the influence of dephasing with the wake when $v_w<c$ in laser-driven cases. The following terms are higher order corrections which account, respectively, for slippage due to $|\bm{v}| < c$ and slippage due to betatron motion.

\begin{figure}[!htb]
	\centering
	\includegraphics*[width=430pt]{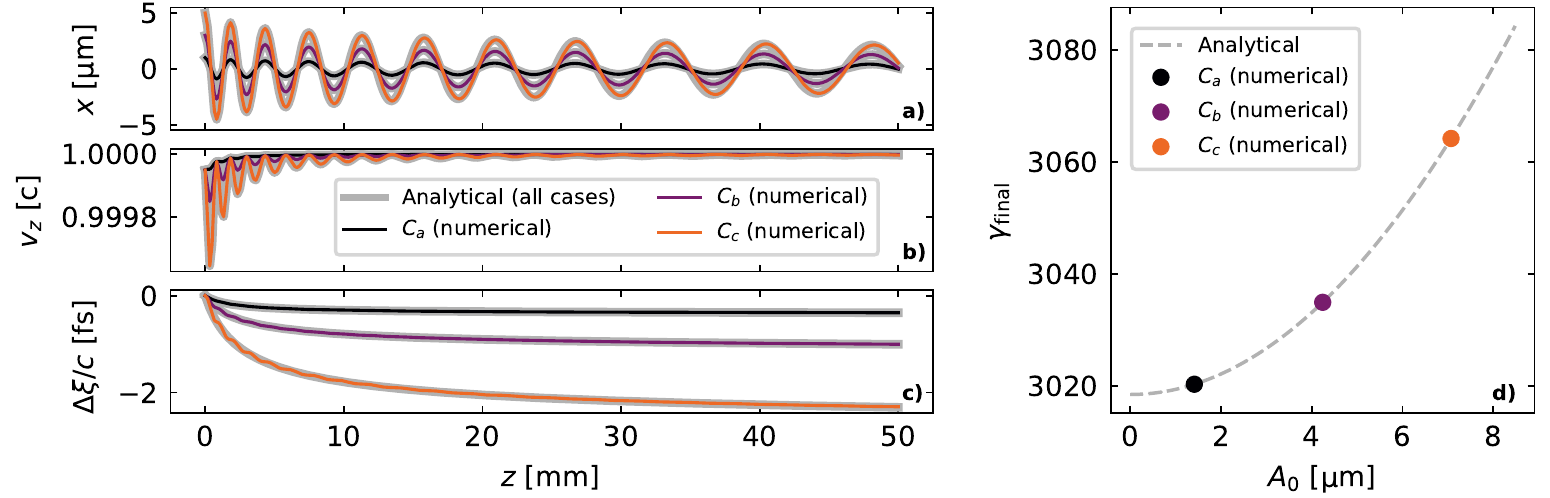}
	\caption{Comparison of the analytical and numerical results for three test particles with different oscillation amplitude in a 5 cm plasma stage: \textbf{(a)} betatron oscillations in $x$, \textbf{(b)} longitudinal velocity, \textbf{(c)} longitudinal slippage, and \textbf{(d)} the final energy after acceleration. The analytical predictions are all shown in grey, while the numerical solutions have a different color for each particle.} 
	\label{fig:model_validation}
\end{figure}

Eqs. (\ref{eq:x}), (\ref{eq:vx}), (\ref{eq:finalXi}) and (\ref{eq:finalGamma2}) allow for a complete description of the single-particle evolution within the wakefields. In order to test their validity, they have been compared with numerical solutions of Eqs. (\ref{eq:lon_eq}) and (\ref{eq:transv_eq}). Three test cases corresponding to single electrons with different initial transverse offsets are shown in Fig. \ref{fig:model_validation}. The electrons in these three cases ($C_a$, $C_b$ and $C_c$) have, respectively, $x_0=y_0= 1, 3, 5 \ \mathrm{\upmu m}$ and $v_{x,0}=v_{y,0}=0$. They are injected with $\gamma_0 = 100$ and propagate 5 cm within a plasma stage with $n_p=10^{17} \ \mathrm{cm^{-3}}$ assuming a blowout with $\mathcal{E}'=-\omega_p^2/2c$, $\mathcal{E}_0 = \omega_p$ and $\mathcal{K}=\omega_p^2/2$. The wake velocity is directly determined from the group velocity of a laser driver as $v_w/c=\omega_l/(\omega_p^2+\omega_l^2)^{1/2}$ if etching effects\cite{PhysRevSTAB.10.061301} are neglected, where $\omega_l=2\pi c/\lambda_l$, assuming a wavelength $\lambda_l=800 \ \mathrm{nm}$. The agreement between numerical and analytical solutions is excellent, showing that this single-particle model accurately describes the coupling between longitudinal and transverse dynamics. From these test cases it can also be seen that betatron oscillations with micron-level amplitude can lead to femtosecond-level slippage and energy differences on the percent range.

When considering a particle bunch, these differences in slippage and energy gain between single particles will have an impact on parameters such as the bunch length and energy spread. One immediate consequence will be an increase of the bunch length, which will in turn also lead to increased correlated energy spread if $\mathcal{E}'\neq0$. Another implication, more critical for FEL applications, will be an intra-bunch slice mixing which will lead to a growth of the slice  energy spread.

\subsection*{Impact on slice energy spread}
In order to obtain an expression for the induced slice energy spread, Eqs. (\ref{eq:finalXi}) and (\ref{eq:finalGamma2}) can be used to calculate the energy difference $\Delta \gamma$ between two beam particles, $p_1$ and $p_2$, with the same $\gamma_0$ but different oscillation amplitude, $A_{0,p_1} = 0$ and $A_{0,p_2} \neq 0$, which at some time $t$ are in the same infinitesimally long beam slice at $\xi_s$ due to the experienced slippage. If $\mathcal{E}'$ and $\mathcal{K}$ are constant along the bunch, then $\mathcal{E}_{0,p_i} = \mathcal{E}_{0,s} - \mathcal{E}' \Delta\xi_{p_i}$ for $i=1,2$, where $\mathcal{E}_{0,s}$ is measured at $\xi_s$. This allows the particle energy evolution, $\gamma_{p_i}(t)$, to be written in terms of the fields at the current slice. Additionally, if $\mathcal{E}' \Delta\xi_{p_i}(t)/\mathcal{E}_{0,s} \ll 1$, then $\mathcal{E}_{0,p_i} \simeq \mathcal{E}_{0,s}$ can be assumed for the higher order terms in $\gamma_{p_i}(t)$ while maintaining the full expression in the leading linear term. This also means that $\Delta\xi_{p_2}(t) - \Delta\xi_{p_1}(t) \simeq A_{0,p_2}^2 \mathcal{K} (\Gamma_s^{-1/2}(t)-1)/ 2c\mathcal{E}_{0,s}$, which, added to the previous considerations, leads to
\begin{equation}\label{eq:deltaGammaSlice}
\Delta \gamma(t) = \gamma_{p_2}(t) - \gamma_{p_1}(t) \simeq 
\frac{\gamma_{0}\mathcal{E}'\mathcal{K} A_{0,p_2}^2 }{2 c\mathcal{E}_{0,s}^2} 
\frac{\left(\Gamma_s(t)^{1/2} - 1\right)^2 }{\Gamma_s(t)^{1/2}} ,
\end{equation}
where $\Gamma_s(t) = 1 + \mathcal{E}_{0,s}t/\gamma_{0}$. The correlation $\Delta \gamma \propto -A_0^2$ (if $\mathcal{E}'<0$) within a beam slice arises because, due to slippage, particles with a higher $A_0$ originally come from positions ahead of their current slice. Since $\mathcal{E}'<0$, the accelerating fields they have experienced as they slip towards the back are smaller than at their current position, thus leading to a lower net energy gain. Eq. (\ref{eq:deltaGammaSlice}) therefore shows that although a positive correlation  $\Delta \gamma \propto A_0^2$ exists for particles \textit{starting} at the same slice, as given by Eq. (\ref{eq:finalGamma2}), this correlation is actually negative when looking at particles which have \textit{ended} in the same slice due to slippage. The induced relative slice energy spread due to this effect can be found from the standard deviation of $\Delta \gamma/\bar{\gamma}_s$ if the mean slice energy is assumed to evolve as $\bar{\gamma}_s(t) \simeq \bar{\gamma}_{0,s} + \mathcal{E}_{0,s}t$, from what it is found that
\begin{equation}\label{eq:sigmaSlice}
\frac{\sigma_{\gamma_s}^{\Delta\xi}}{ \bar{\gamma}_s }(t) \simeq 
\frac{\mathcal{E}'\mathcal{K}\sigma_{A^2}}{2 c \mathcal{E}_0^2} 
\frac{\left(\bar{\Gamma}_s(t)^{1/2} - 1\right)^2}{\bar{\Gamma}_s(t)^{3/2}}
\simeq 
\sqrt{\frac{2\mathcal{K}}{\bar{\gamma}_{0,s}}}\frac{\mathcal{E}'\mathcal{F}_m\epsilon_n}{\mathcal{E}_0^2} 
\frac{\left(\bar{\Gamma}_s(t)^{1/2} - 1\right)^2}{\bar{\Gamma}_s(t)^{3/2}} \ ,
\end{equation}
where $\bar{\Gamma}_s = \bar{\gamma}_s/\bar{\gamma}_{0,s}$, $\sigma_{A^2}$ is the standard deviation of $A_0^2$ for the particles within the slice and $\mathcal{E}_0$ is taken at the slice position. For an on-axis Gaussian bunch with zero average transverse momentum ($\bar{x}=\bar{y}=0$ and $\bar{p}_x=\bar{p}_y=0$), it can be obtained that $\sigma_{A^2} \simeq \sqrt{8c^2\mathcal{A}^2/ \mathcal{K} \bar{\gamma}_{0,s}}$, where $\mathcal{A}^2 = \frac{1}{4}\sum_{i=x,y}\epsilon_{n,i}^2\left(\beta_i^4 +\beta_m^4\right)/\beta_i^2\beta_m^2$. The variables $\epsilon_{n,i}$, $\beta_{i} = \sigma_{i}^2\bar{\gamma_{0}}/\epsilon_{n,i}$ and $\sigma_{i}$ are, respectively, the beam's normalized emittance \cite{floettmann2003some}, beta function and rms size in both transverse planes, while $\beta_m = c/\omega_0$, with  $\omega_0=\omega_{x,0}=\omega_{y,0}$, is the matched beta function \cite{assmann1998transverse} defined by the focusing fields within the plasma. For a cylindrically symmetric bunch, i.e, if $\epsilon_{n,i}=\epsilon_{n}$ and $\beta_i=\beta$, a "mismatch factor" $\mathcal{F}_m=((M^4+1)/2M^2)^{1/2}$ can be defined, where $M=\beta/\beta_m$, such that $\mathcal{A}=\mathcal{F}_m\epsilon_n$, which leads to the last expression in Eq. (\ref{eq:sigmaSlice}). For a matched beam ($M=1$) the mismatch factor is $\mathcal{F}_m=1$, which also corresponds to its minimum value. The impact of betatron slippage on the slice energy spread will therefore be stronger on mismatched beams. 

Analyzing Eq. (\ref{eq:sigmaSlice}) it can be seen that the induced slice energy spread has a fast initial growth until it reaches a maximum when the bunch energy has increased by a factor 9, i.e, when $\bar{\Gamma}_s=9$. For higher energies, the slice energy spread slowly decreases proportional to $\bar{\Gamma}_s^{-1/2}$. This behaviour arises from $\sigma_{\gamma_s}^{\Delta\xi}$, whose growth rate is initially faster than $\bar{\gamma}_s$ but gradually slows down due to the reduced slippage at higher energies. 

\begin{figure}[!htb]
	\centering
	\includegraphics*[width=300pt]{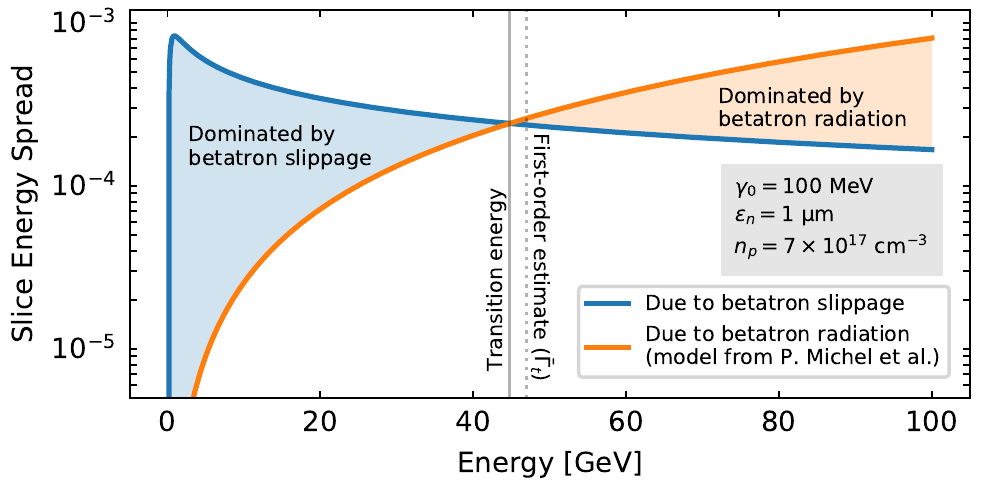}
	\caption{Comparison of the slice energy spread due to slippage (from Eq. (\ref{eq:sigmaSlice})) and betatron radiation (from model by P. Michel \textit{et al.} \cite{michel2006radiative}) for the particular case of a 100 MeV bunch with 1 mm mrad normalized emittance injected with matched beta function into a plasma with $n_p=7 \times 10^{17} \ \mathrm{cm^{-3}}$ assuming a wakefield in the blowout regime ($\mathcal{E}'=-\omega_p^2/2c$, $\mathcal{E}_0 = \omega_p$, $\mathcal{K}=\omega_p^2/2$). Betatron slippage is found to be the dominant effect up to an energy of 44.8 GeV, which is accurately estimated by the $\sim47 \ \mathrm{GeV}$ value given by the first order approximation $\bar{\Gamma}_t \simeq \mathcal{R}^{-1/2}$.}
	\label{fig:slip_rad_comp}
\end{figure}

This source of energy spread will act in addition to that due to betatron radiation (BR) \cite{wang2002x,esarey2002synchrotron}, which has also been shown to induce a correlation $\Delta \gamma \propto -A_0^2$ and generate a slice energy spread given by \cite{michel2006radiative}
$\sigma_{\gamma_s}^{\mathrm{BR}}/\bar{\gamma}_s \simeq C_{\mathrm{BR}} (\bar{\Gamma}_s^{5/2} -1)/\bar{\Gamma}_s$, where $C_{\mathrm{BR}} = (2 r_e/15c^3)\mathcal{K}^2\sigma_{A^2}\bar{\gamma}_{0,s}^2/\mathcal{E}_0$ and $r_e=e^2/4\pi\epsilon_0mc^2$ is the classical electron radius. As both effects are induced by the same underlying cause (the betatron motion of particles), it is of interest to compare and determine their relative relevance. This can be done by obtaining the intersection points of Eq.  (\ref{eq:sigmaSlice}) with the BR expression, for which perturbation theory is needed. This method yields that the slippage-induced energy spread will initially dominate over BR if the ratio $\mathcal{R}=C_{\mathrm{BR}}/C_{\mathrm{\Delta\xi}} \ll 0.1$, where $C_{\mathrm{\Delta\xi}} = \mathcal{E}'\mathcal{K}\sigma_{A^2}/2 c\mathcal{E}_0^2 $ is the coefficient in Eq. (\ref{eq:sigmaSlice}). When this is the case, a transition into a BR-dominated regime occurs at an energy $\bar{\Gamma}_s=\bar{\Gamma}_t$ which, to first order, can be obtained as $\bar{\Gamma}_t \simeq \mathcal{R}^{-1/2}$. An illustration of both regimes can be seen in Fig. \ref{fig:slip_rad_comp} for a particular set of parameters. As a general rule, although this will vary from case to case, the energy spread due to slippage will typically dominate over BR up to energies on the order of $\sim 10$ GeV. This is the energy range in which FELs operate, implying that betatron slippage is the source of slice energy spread that should be minimized for these applications.

An effective way for mitigating the slippage-induced energy spread would be to beam-load the wake such that $\mathcal{E}'\to0$, thus suppressing the energy spread growth as given by Eq. \ref{eq:sigmaSlice}. However, since this condition is difficult to achieve, especially along the whole bunch, other strategies can be derived from this equation. Regarding the accelerating structures, possible strategies include reducing plasma density, since the ratio $\mathcal{K}^{1/2}\mathcal{E}'/\mathcal{E}_0^2\propto n_p^{1/2}$, or using hollow plasma channels \cite{chiou1995laser}, so that $\mathcal{K}\to 0$. In regards to the accelerated beam, minimizing the emittance, increasing the initial energy and fulfilling matching conditions ($\mathcal{F}_m =1$) would be beneficial, with these two last points being especially relevant for externally injected beams.

Minimizing this slice energy spread is especially important for PBAs based on chirp compensation schemes \cite{manahan2017single, brinkmann2017chirp, PhysRevLett.123.054801}. This is because although the average $\mathcal{E}'$ experienced by the beam after acceleration might be 0, its value at any point of the accelerator is not, thus giving rise to the development of a slice energy spread which will remain in the de-chirped beam. It is possible, however, that the alternation between positive and negative $\mathcal{E}'$ in these schemes might partially mitigate the correlation $\Delta\gamma\propto A_0^2$ from Eq. (\ref{eq:deltaGammaSlice}) and thus the slice energy spread, although never remove it completely. As an additional remark, it should also be noted that this correlation between oscillation amplitude and energy caused by betatron slippage has been proposed as a way of performing beam conditioning for FELs \cite{sessler1992radio} by means of a magnetic chicane \cite{pennt2007plasma}. Thus, it might also be beneficial in some cases.


\subsection*{Impact on bunch length and total energy spread of ultra-short bunches}
In addition to its impact on the slice energy spread, another immediate consequence of the betatron-induced slippage will be a growth of the bunch length. This is especially relevant for ultra-short bunches where the slippage experienced by the particles is comparable to or bigger than the initial bunch length $\sigma_{z,0}$. The contribution of the single-particle slippage to $\sigma_z$ can be directly obtained from the standard deviation of Eq. (\ref{eq:finalXi})
by assuming an initially monoenergetic bunch with $\sigma_{z,0} =0$. The bunch lengthening is then found to be
\begin{equation}\label{eq:sigmaLengthening}
\sigma_{\Delta\xi}(t) \simeq \frac{\mathcal{K} \sigma_{A^2} }{2c\mathcal{E}_0} 
\left(1-\bar{\Gamma}(t)^{-1/2}\right)
\simeq
\sqrt{\frac{2\mathcal{K}}{\bar{\gamma}_{0}}}\frac{\mathcal{F}_m\epsilon_n}{\mathcal{E}_0}\left(1-\bar{\Gamma}(t)^{-1/2}\right)
 ,
\end{equation}
where $\bar{\Gamma} = \bar{\gamma}/\bar{\gamma}_{0} \simeq 1 + \mathcal{E}_{0}t/\bar{\gamma}_{0}$. The last expression on the right is again obtained for the case of a cylindrically-symmetric Gaussian bunch. The bunch length evolution, when taking into account the impact of slippage, is thus given by $\sigma_z(t) = (\sigma_{z,0}^2 + \sigma_{\Delta\xi}^2(t))^{1/2}$ for initially Gaussian bunches. When this is the case, this expression therefore establishes a limitation on the minimum bunch length achievable in a PBA. This could have a strong impact on the production of high-energy, sub-femtosecond bunches, because although certain techniques to inject such ultra-short bunches into the plasma wake have been proposed\cite{li2013dense, WEIKUM201633, PhysRevLett.119.044801, doi:10.1063/1.4929921}, their sub-femtosecond duration could be lost when further accelerating them. Possible ways of mitigating this lengthening can be found by looking at Eq. (\ref{eq:sigmaLengthening}), which yields the same conditions on the beam emittance, matching and initial energy found for Eq. (\ref{eq:sigmaSlice}). In this case, however, no benefits from beam-loading or reduced plasma density are expected because $\mathcal{E}'$ does not play a role in the lengthening and the ratio $\mathcal{K}^{1/2}/\mathcal{E}_0$ does not depend on the plasma density. Another remark that can be extracted from Eq. (\ref{eq:sigmaLengthening}) is that, as expected from Eq. (\ref{eq:xiLim}), the lengthening has a finite upper limit when $\bar{\Gamma} \to \infty$ given by $ \sigma_{\Delta\xi}^\mathrm{max} \simeq (2\mathcal{K}/\bar{\gamma}_{0})^{1/2} \mathcal{F}_m \epsilon_n/\mathcal{E}_0$.

As a consequence of the bunch lengthening, in cases where $\mathcal{E}'\neq0$, the correlated energy spread of the bunch will also increase. This explains previously observed limitations, where convergence to a finite energy spread was found as $\sigma_{z,0} \to 0$ \cite{grebenyuk2014laser}. This limit on the minimum energy spread as the initial bunch length is reduced can be calculated for an initially monoenergetic beam with $\sigma_{z,0} = 0$ from the standard deviation of Eq. (\ref{eq:finalGamma2}). This leads to
\begin{equation}\label{eq:sigmaEneTot}
\frac{\sigma_{\gamma}^{\Delta\xi}}{ \bar{\gamma} }(t) \Bigg|_{\sigma_{z,0} \to 0}\simeq \frac{\mathcal{E}'\mathcal{K}\sigma_{A^2}}{2 c \mathcal{E}_0^2}
\frac{\left(\bar{\Gamma}(t)^{1/2} - 1\right)^2}{\bar{\Gamma}(t)}
\simeq
\sqrt{\frac{2\mathcal{K}}{\bar{\gamma}_{0}}}\frac{\mathcal{E}'\mathcal{F}_m\epsilon_n}{\mathcal{E}_0^2} \frac{\left(\bar{\Gamma}(t)^{1/2} - 1\right)^2}{\bar{\Gamma}(t)} \ ,
\end{equation}
which, in the same way as Eq. (\ref{eq:sigmaLengthening}), also tends to a finite value as $\bar{\Gamma} \to \infty$. The last expression on the right corresponds again to the case of a cylindrically-symmetric Gaussian bunch.

\begin{figure}[!htb]
	\centering
	\includegraphics*[width=450pt]{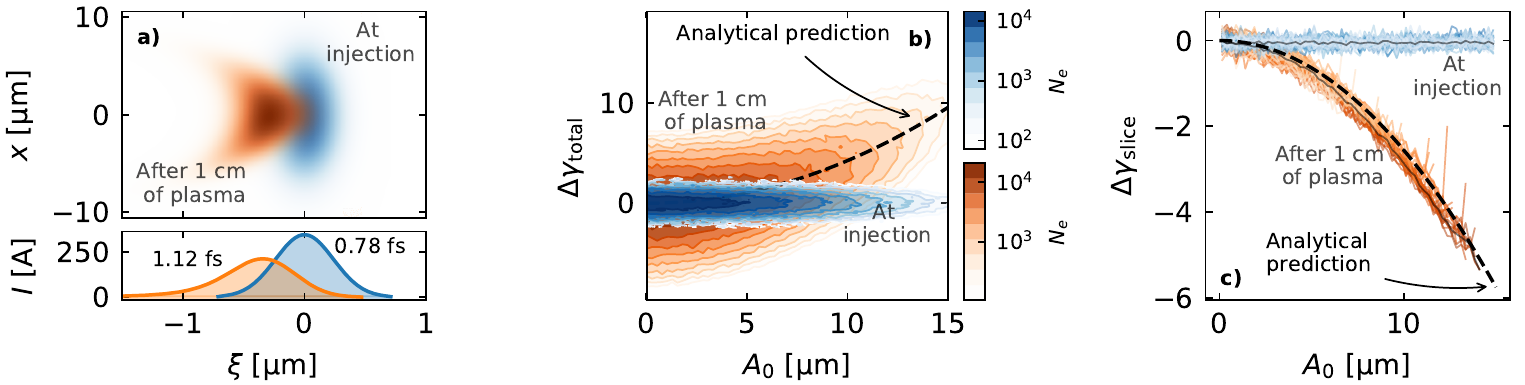}
	\caption{Results from an illustrative 2D OSIRIS simulation of a 100 MeV, 0.7 pC and initially sub-femtosecond Gaussian electron bunch externally injected into the blowout of a laser driver in a plasma with $n_p=10^{17} \ \mathrm{cm^{-3}}$. More details can be seen in the Methods section. \textbf{(a)} Spatial distribution and current profile of the electron beam at injection and after 1 cm of propagation in the plasma. The development of a curved shape and increased length, as expected from the particle slippage, are clearly visible. \textbf{(b)} Beam energy distribution with respect to the initial oscillation amplitude of the particles. As expected from Eq. (\ref{eq:finalGamma2}), those with higher $A_0$ experience a higher energy gain. The analytical prediction of the energy-amplitude correlation as given by this equation is shown by a dashed black line. This curve was computed from the value of the fields in the simulation at the central position of the bunch, i.e., $\mathcal{E}_0\simeq0.60 \, \omega_p$, $\mathcal{E}'\simeq-0.27 \, \omega_p^2/c$ and $\mathcal{K}\simeq0.36 \, \omega_p^2$. The energy difference $\Delta \gamma_\mathrm{total}$ was calculated with respect to the average energy of the particles with $A_0\simeq0 \ \mathrm{\upmu m}$. \textbf{(c)} Distribution of the particle energy with respect to the initial oscillation amplitude within each beam slice. The coloured lines correspond to the distribution within each 20-nm-long slice, while the thin dark lines represent the average. As expected from Eq. (\ref{eq:deltaGammaSlice}), the energy-amplitude correlation at the slice level is negative. The analytical prediction was again computed for the central beam slice using the same field values as before. The energy difference within the slice, $\Delta \gamma_\mathrm{slice}$, was calculated with respect to the average energy of the particles with $A_0\simeq0 \ \mathrm{\upmu m}$ after removing the longitudinal energy correlation (chirp) along the beam.
	}
	\label{fig:showcaseOfEffects}
\end{figure}

An illustrative example of the phenomena described in this section can be seen in Fig. \ref{fig:showcaseOfEffects}, where an externally injected witness beam is shown at the entry and after 1 cm of plasma, as obtained from an OSIRIS 2D simulation. The induced lengthening as well as the development of a quadratic correlation between energy and oscillation amplitude can be clearly seen. As predicted by Eqs. (\ref{eq:finalGamma2}) and (\ref{eq:deltaGammaSlice}), this correlation is positive for the overall bunch (Fig. \ref{fig:showcaseOfEffects}b) but negative on the slice level (Fig. \ref{fig:showcaseOfEffects}c).

\subsection*{Validation against 3D PIC simulations}
In order to validate and illustrate the significance of the expressions for the slice energy spread and bunch lengthening, a series of 3D simulations with the PIC codes OSIRIS and HiPACE have been performed. The studied cases consisted on a beam-driven plasma stage providing a 1 GeV net energy gain to an externally injected witness beam, as it would be interesting for FEL applications or multi-staged acceleration. In the OSIRIS simulations, the impact of the slippage effects has been tested for different witness beam parameters and plasma densities, and in all cases beam loading effects could always be neglected. On the contrary, the HiPACE simulations where performed to study the possible mitigating effects of beam loading for a fixed set of plasma and beam parameters, where only the total beam charge was varied. Although only beam-driven cases have been simulated, similar results may apply for a laser driver as long as dephasing effects in the mean beam energy can be neglected, as they are not included in Eqs. (\ref{eq:sigmaSlice}) to (\ref{eq:sigmaEneTot}).

In all simulated cases, a Gaussian driver with an energy of 1 GeV and a $0.1 \%$ spread has been considered. Its dimensions and charge have been defined in terms of the plasma density so that the generated blowout can be scaled with $n_p$. For the baseline case in which $n_p=7 \times 10^{17} \ \mathrm{cm^{-3}}$, as used in previous experiments \cite{leemans2014multi}, this means a transverse size $\sigma_{x}=\sigma_{y} = 0.4 \, c/\omega_p \simeq 2.5 \ \mathrm{\upmu m}$, a length $\sigma_z = c/\omega_p \simeq 6.4 \ \mathrm{\upmu m}$, a normalized emittance $\epsilon_{n} = 0.4 \, \sigma_x \simeq 1 \ \mathrm{\upmu m \ rad}$ and a peak electron density $n_b\simeq3.78 \, n_p$ for a total charge $Q\simeq265$ pC and a peak current $I_p \simeq 5 \ \mathrm{kA}$. The plasma target has a flat-top longitudinal profile and is transversely uniform.

For the studies made with OSIRIS, a Gaussian witness beam is injected on-axis at $\xi_w \simeq \xi_d - 4.5 \, c/\omega_p$, where $\xi_d$ is the driver centre. Different emittance values ($\epsilon_{n}= 0.1, \ 1 \ \mathrm{and} \ 10 \ \mathrm{\upmu m \, rad}$) as well as injection energies between 10 MeV and 1 GeV have been tested. The transverse size is matched to the focusing fields so that $\mathcal{F}_m=1$. With regards to the bunch charge and duration, two different sets of parameters have been considered: $\sigma_z/c = 3 \ \mathrm{fs}$ with 1 pC for the studies on the slice energy spread, as it could be obtained in the SINBAD facility at DESY \cite{assmann2014sinbad, zhu2016sub}, while a shorter bunch with $\sigma_z/c = 0.1 \ \mathrm{fs}$ with 0.1 pC was used for the bunch lengthening studies. This shorter bunch duration was chosen to better highlight the impact of lengthening due to slippage on sub-femtosecond bunches. The peak current in both cases is too low to significantly modify $\mathcal{E}'$ along the bunch, therefore allowing beam loading effects to be neglected. The slice energy spread in all tested beams is dominated by slippage, as $\mathcal{R}\ll0.1$ and $\bar{\Gamma} \ll \mathcal{R}^{-1/2}$. Also, in order to isolate the plasma stage as the only source of energy spread, a zero longitudinal momentum spread at injection has been considered.

\begin{figure}[!htb]
	\centering
	\includegraphics*[width=480pt]{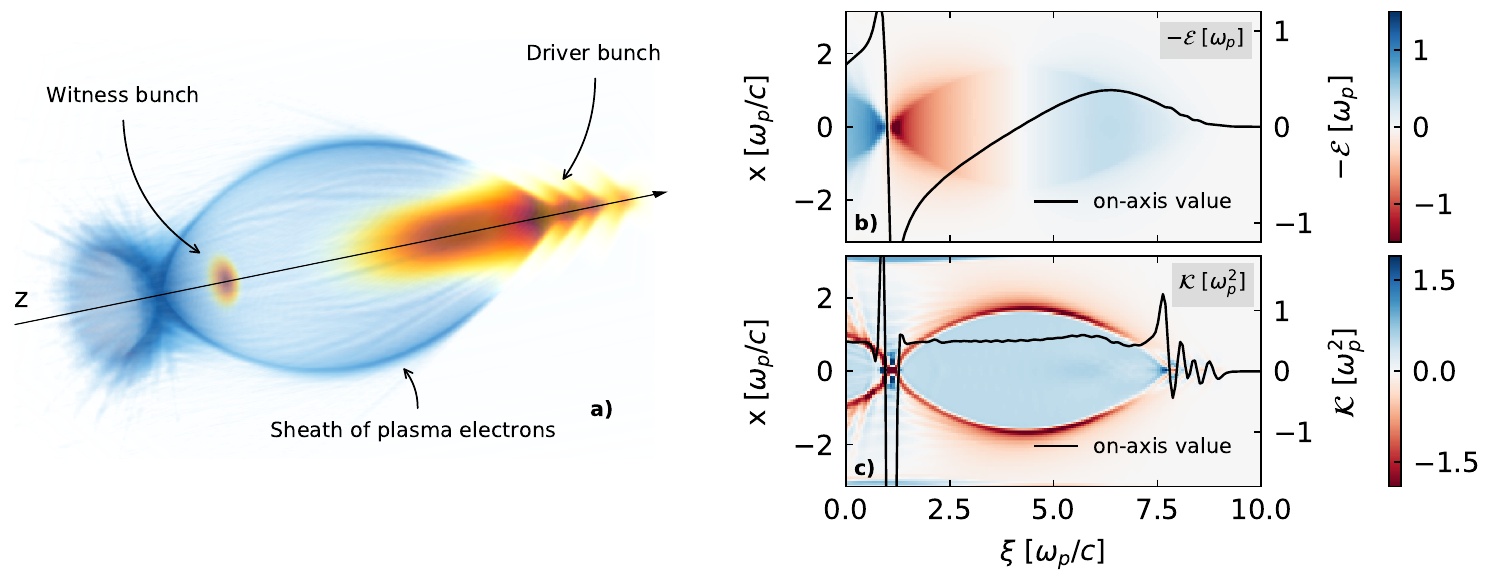}
	\caption{\textbf{(a)} 3D visualization made with VisualPIC \cite{FerranPousa:2017umq} of one of the OSIRIS simulations. The case shown is that of a witness beam with $\epsilon_n=10 \ \mathrm{\upmu m \, rad}$ and $n_p=10^{17} \ \mathrm{cm^{-3}}$, taken at a time $t\simeq816/\omega_p$. \textbf{(b)} and \textbf{(c)} show the longitudinal field $\mathcal{E}$ and focusing gradient $\mathcal{K}$ of the same simulated case along the central slice ($y=0$) and on axis (black line).}
	\label{fig:field_plots}
\end{figure}

In order to compare the simulation results with the analytical theory, the field values in Eqs. (\ref{eq:sigmaSlice}) and (\ref{eq:sigmaLengthening}) need to be provided. While these could be estimated from analytical models \cite{lu2006nonlinear,yi2013analytic}, they were measured directly at $\xi_w$ from a single simulation with no witness beam, obtaining $\mathcal{E}_0\simeq0.53 \, \omega_p$, $\mathcal{E}'\simeq-0.37 \, \omega_p^2/c$ and $\mathcal{K}\simeq0.5 \, \omega_p^2$. This allows for a more accurate validation of the model. A view of the plasma wakefields in one of the OSIRIS simulations is shown in Fig. \ref{fig:field_plots}. As assumed by the analytical model, a constant $\mathcal{K}$ within the blowout and a transversely homogeneous $\mathcal{E}$ with uniform $\mathcal{E}'$ along the witness bunch can be seen.

\begin{figure}[!htb]
	\centering
	\includegraphics*[width=310pt]{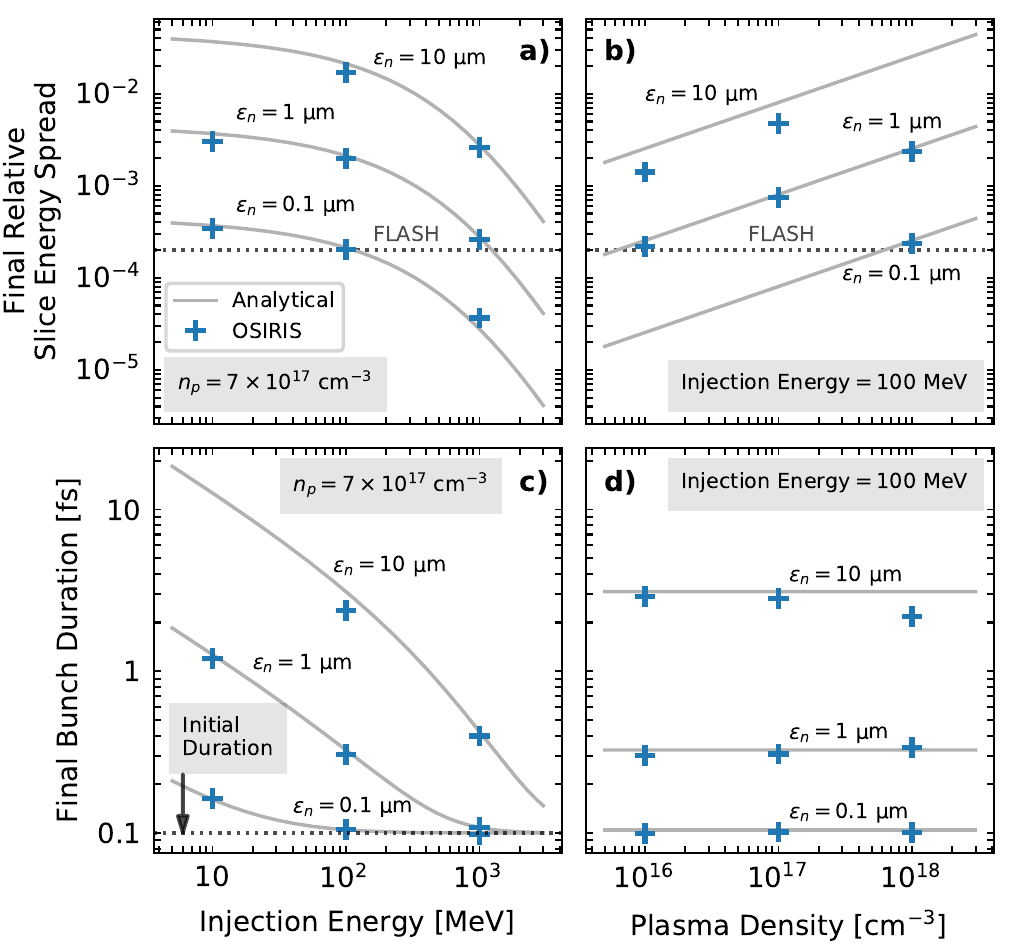}
	\caption{Comparison between simulation results and the analytical curves from Eqs. (\ref{eq:sigmaSlice}) and (\ref{eq:sigmaLengthening}) for a 1 GeV plasma stage, corresponding to an acceleration length of $\sim3240 \omega_p/c$ in all cases. The influence of injection energy, \textbf{(a)} and \textbf{(c)}, and plasma density, \textbf{(b)} and \textbf{(d)}, on the generated slice energy spread and bunch lengthening is displayed. The energy spread of the FLASH FEL at DESY ($\sim 2 \times 10^{-4}$) \cite{schreiber2015free} is shown for reference. Only results from simulations with no significant numerical noise and in which the witness beam was sufficiently narrow to fit within the wake have been included.
	}
	\label{fig:beamLength}
\end{figure}

\begin{figure}[!htb]
	\centering
	\includegraphics*[width=500pt]{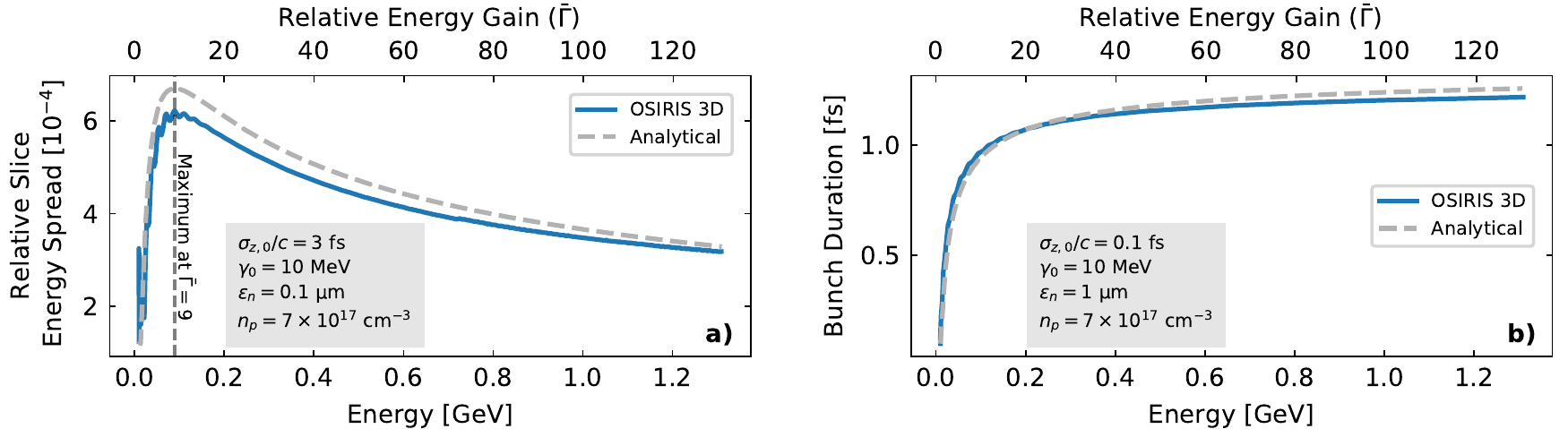}
	\caption{Evolution of the relative slice energy spread \textbf{(a)} and bunch duration \textbf{(b)} on two of the simulated cases. Excellent agreement between analytical model and simulations is shown, including the predicted energy spread maximum at $\bar{\Gamma}=9$. }
	\label{fig:single_case}
\end{figure}

The final simulation results, displayed in Fig. \ref{fig:beamLength}, show a strong dependence of the generated slice energy spread on the plasma density and initial beam parameters, reaching values in excess of $10^{-2}$ (or 1\%) for certain configurations, way above FEL requirements. Similarly, the increase in bunch length can be on the femtosecond level for typical emittance and energy values, showcasing the difficulty of achieving GeV-class, sub-femtosecond bunches. No variation of the bunch length with the plasma density is observed, as expected from Eq. (\ref{eq:sigmaLengthening}). The analytical model shows excellent agreement with the simulations over several orders of magnitude, with expected discrepancies in some of the higher emittance cases. The differences in Fig. \ref{fig:beamLength}(b) arise from the large slippage experienced by particles with higher oscillation amplitude, which causes the transverse distribution of many beam slices to be a truncated Gaussian and thus leads to a smaller energy spread than predicted. The discrepancies in Figs. \ref{fig:beamLength}(c) and (d) 
arise from the longitudinal energy correlation, neglected by Eq. (\ref{eq:sigmaLengthening}), which induces a slight longitudinal bunch compression due to velocity differences between head and tail. In addition to the final slice energy spread and bunch length values, the analytical model also accurately predicts the evolution of these parameters during propagation within the PBA, as shown in Fig. \ref{fig:single_case}. The discrepancies in the energy spread evolution in Fig. \ref{fig:single_case}(a) arise from the initial dephasing between the bunch and plasma wakefield due to the low initial energy (10 MeV), which leads to an experienced accelerating field $\mathcal{E}>\mathcal{E}_0$ and, thus, a lower energy spread than predicted by Eq. (\ref{eq:sigmaSlice}). Furthermore, the initial oscillations on the slice energy spread at the beginning of the simulation correspond to the betatron oscillations of the witness beam, which lead to an oscillation in $v_z$, as seen previously in Fig. \ref{fig:model_validation}(b). When the $v_z$ of the beam particles is minimum (maximum), more (less) slippage occurs and the energy spread growth is faster (slower). This also leads to small oscillations on the bunch length, as seen in Fig. \ref{fig:single_case}(b). The impact of the $v_z$ oscillations on the particle slippage had to be neglected in the theoretical model in order to obtain the analytical expression in Eq. (\ref{eq:finalXi}). Thus, this effect is not reproduced by the analytical curves shown in Fig. \ref{fig:single_case}.

\begin{figure}[!htb]
	\centering
	\includegraphics*[width=500pt]{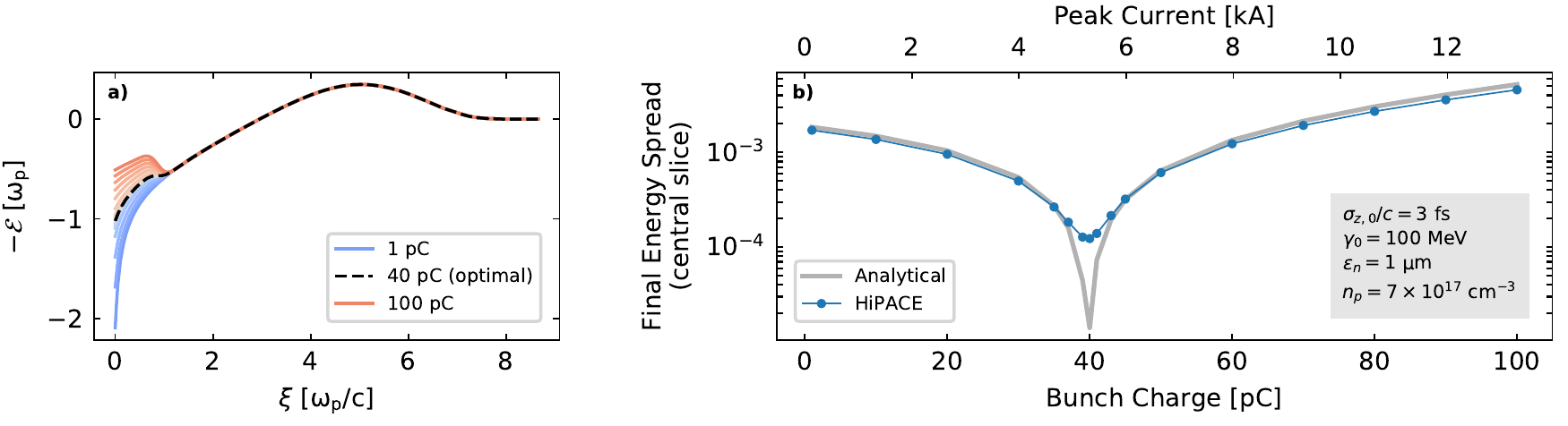}
	\caption{\textbf{(a)} Effect of beam loading on the on-axis longitudinal electric field in all simulated cases for different total bunch charge. The ideal case which minimizes $\mathcal{E}'$ is shown in black, while those in which $\mathcal{E}'<0$ are shown in blue tones and those in which beam loading inverts the slope ($\mathcal{E}'>0$) are displayed in red. The fields are shown at $t=500/\omega_p$. \textbf{(b)} Final central slice energy spread of the bunch with respect to its total charge as obtained from the HiPACE simulations. The value of the energy spread is calculated after a 1 GeV energy gain in the plasma. The effect of beam loading is clear, providing a minimum energy spread for a charge of 40 pC, which corresponds to a peak current of $\sim5.3$ kA. }
	\label{fig:beam_loading}
\end{figure}

In order to explore the potential of beam loading for mitigating the generated slice energy spread, an additional set of simulations was performed with HiPACE. The same driver parameters as in the OSIRIS simulations and a plasma density of $7\times 10^{17} \ \mathrm{cm^{-3}}$ were used. The witness beam parameters were fixed to an initial energy of 100 MeV, $\epsilon_n = 1 \ \mathrm{\upmu m \, rad}$ and $\sigma_z/c = 3 \ \mathrm{fs}$, while its total charge was varied from 1 to 100 pC. The final energy spread of the central bunch slice after a 1 GeV energy gain is shown in Fig. \ref{fig:beam_loading} for the different bunch charges. Only the central slice energy spread is displayed in this case because the effect of beam loading on $\mathcal{E}$ varies along the bunch. The analytical points were calculated from the average value during acceleration of the fields at the center of the bunch. 

The results show again good agreement between theory and simulations, and that beam loading can significantly reduce the development of slice energy spread (up to a factor of $\sim 20$ in this case). For the particular set of parameters under consideration, the energy spread minimum is achieved for a total bunch charge of 40 pC (corresponding to a peak current of $\sim5.3$ kA) and grows again for higher charges because beam loading overcompensates $\mathcal{E}'$, which becomes positive. Peak currents in this range of parameters have been experimentally demonstrated \cite{lundh2011few,couperus2017demonstration} and, if optimized, could therefore provide an effective way of mitigating the slice energy spread growth. Discrepancies between the analytical curve and the simulations appear, however, around the energy spread minimum. This is because, in theory, the slice energy spread due to slippage could be completely suppressed for an ideally beam-loaded case in which $\mathcal{E}'=0$ along the whole bunch. This, however, requires an optimized trapezoidal current profile \cite{tzoufras2008beam} , while the bunch in the simulated case is Gaussian and therefore does not experience a uniform $\mathcal{E}'$. Thus, even if $\mathcal{E}'=0$ at the center, a certain energy spread will develop in the central slice due to the slippage of particles originally coming from slices ahead, where $\mathcal{E}'\neq0$. In addition, other effects apart from particle slippage can lead to increased slice energy spread in beam-loaded cases, such as transverse variations of the accelerating fields in the linear and weakly non-linear regimes \cite{PhysRevAccelBeams.21.111301}.

\section*{Discussion}

The presented analytical model and simulation results show that the impact of the betatron-induced slippage could be very significant on beam parameters such as the slice energy spread (up to $\sim1\%$) and bunch length (growth on the femtosecond range). These effects should therefore be minimized, and guidelines for their mitigation can be derived from the analytical expressions together with the simulated cases. These show that an effective method of suppressing the slippage-induced slice energy spread growth is the use of beam loading (so that $\mathcal{E}' \to 0$), while another possible solution would be performing the acceleration in hollow plasma channels ($\mathcal{K} \to 0$). Otherwise, in order not to rely on these two options, the simulations show that for achieving a slice energy spread $<10^{-3}$, as typically required by FELs, a plasma density $\lesssim 10^{17} \ \mathrm{cm^{-3}}$, a normalized emittance $\lesssim 1 \ \mathrm{\upmu m \, rad}$ as well as an initial energy $\gtrsim 100 \ \mathrm{MeV}$ become necessary. The same considerations apply regarding the production of sub-femtosecond bunches, although beam loading or lower plasma densities do not provide any benefit in this case. For applications requiring multiple accelerating stages, such as plasma-based particle colliders, these results imply that the slice energy spread due to slippage would mostly be generated in the first accelerating stage, decreasing at higher energies and eventually being dominated by the emission of betatron radiation.

\section*{Methods}

\subsection*{3D Particle-in-Cell simulations with OSIRIS}
The 3D PIC in simulations corresponding to the results shown in Figs. \ref{fig:field_plots}, \ref{fig:beamLength} and \ref{fig:single_case} where carried out with the OSIRIS code \cite{fonseca2002osiris} using a standard finite differences field solver \cite{1138693}. Due to the long propagation distance within the plasma, a co-moving simulation box was considered. The simulation grid on the cases shown in Figs. \ref{fig:beamLength} and \ref{fig:single_case} consisted on $146\times106\times106$ cells and had dimensions of $8.66\times6.30\times6.30$ in units of $\omega_p/c$. In the graphs shown in Fig. \ref{fig:field_plots} used for visualizing the fields, the simulation grid was slightly longer, with  $170\times106\times106$ cells and dimensions of $10\times6.30\times6.30$ in units of $\omega_p/c$. This was done in order to show the tail of the wakefields, which was otherwise cut off. The time step in all cases was chosen to fulfil the stability criterion \cite{1138693}. The number of particles per cell was 16 for the beam driver, 1 for the plasma and between 64 and 4096 for the witness beam, depending on the case, in order to have a meaningful total number of particles. The initialization of the electron beams was done using the \textit{n\_accelerate} method in OSIRIS, which progressively increases the longitudinal particle momentum over the specified number of time steps in order to obtain almost self-consistent electromagnetic fields before the beam enters the plasma. 

\subsection*{3D Particle-in-Cell simulations with HiPACE}
The 3D PIC in simulations corresponding to the results shown in Fig. \ref{fig:beam_loading} where carried out with the HiPACE code \cite{Mehrling_2014}, which is based on the quasi-static approximation. The simulation grid consisted on $320\times106\times106$ cells and had dimensions of $8.66\times6.30\times6.30$ in units of $\omega_p/c$. A constant time step size of 5 was used. The number of particles per cell was 16 for the beam driver, 3 for the plasma and 512 for the witness beam in order to have a meaningful total number of particles. The initialization of the particle species in HiPACE is performed by computing the self-consistent fields in each time step by means of fast Poisson solvers.

\subsection*{2D Particle-in-Cell simulation with OSIRIS}
The data displayed in Fig. \ref{fig:showcaseOfEffects} showcasing the bunch lengthening and energy spread development comes from a single 2D simulation with OSIRIS of a laser-driven PBA. This simulation considered a plasma density $n_p=10^{17} \ \mathrm{cm^{-3}}$, a Gaussian laser pulse with a peak power of 1.5 PW and a total energy of 25 J, a peak normalized vector potential $a_0=4$, a wavelength $\lambda=800$ nm, a spot size $w_0=54 \ \mathrm{\upmu m}$ and a FWHM duration $\tau=15 \ \mathrm{fs}$. The witness bunch has a Gaussian profile based on achievable parameters by the ARES linac at SINBAD \cite{assmann2014sinbad, zhu2016sub}, featuring a total charge of 0.7 pC, an initial energy of 100 MeV, a duration $\sigma_z/c=0.78 \ \mathrm{fs}$, a transverse size with $\sigma_x=5.0 \ \mathrm{\upmu m}$ and $\sigma_y=5.3 \ \mathrm{\upmu m}$, an emittance $\epsilon_x=0.70 \ \mathrm{\upmu m \,rad}$ and $\epsilon_y=0.74 \ \mathrm{\upmu m \,rad}$ and an energy spread $\sigma_\gamma/\bar{\gamma}=0.37 \%$. This simulation was also performed with a standard finite differences field solver \cite{1138693} considering a co-moving simulation box with $5000\times500$ cells and dimensions of $5.95\times29.76$ in units of $\omega_p/c$. The time step was chosen to fulfil the stability criterion \cite{1138693}. The number of particles per cell was 8 for the plasma and 64 for the witness beam. The initialization method of the witness beam is the same as in the 3D case described above.

\subsection*{Numerical solution of the single-particle equations of motion}
As shown in Fig. \ref{fig:model_validation}, the equations of motion for a single particle (Eqs. (\ref{eq:lon_eq}) and (\ref{eq:transv_eq})) were solved numerically to compare against the analytical expressions. These equations were solved with a Runge-Kutta method of order 4 with a time step of 0.01 fs. The fields experienced by the particles take into account dephasing effects assuming a laser driver with $\lambda=800$ nm.

\subsection*{Calculation of the slice energy spread}
Due to the limited amount of particles in a PIC simulation, calculating $\sigma_{\gamma}^s / \bar{\gamma}$  for an infinitesimally short slice as in Eq. (\ref{eq:sigmaSlice}) is not possible. However, this quantity can be estimated for a finite slice in which the longitudinal energy correlation is removed. The uncorrelated energy distribution can then be obtained as $\gamma_\mathrm{unc} = \gamma - \delta \left(\xi - \bar{\xi}\right)$, where $\delta$ is the slope of the correlation. Additionally, since $\delta$ might change along the bunch, the bunches are divided into $n$ slices in which $\sigma_{\gamma}^s / \bar{\gamma}$ is calculated. From there, for the cases shown in Figs. \ref{fig:beamLength} and \ref{fig:single_case}, the slice energy spread of the whole bunch is obtained as the weighted average of the value on each slice. This process is repeated by increasing the number of slices until convergence to a value is reached while still having a significant amount of particles per slice. For the cases to study the effect of beam loading shown in Fig. \ref{fig:beam_loading}, the value of the slice energy spread only given for the central slice and not as an average.

\section*{Data availability}
All data displayed in the included figures is available at \url{https://bib-pubdb1.desy.de/record/415560/}. Further data supporting this study is available from the corresponding author upon reasonable request.

\bibliography{intrinsic_growth}

\section*{Acknowledgements}

This project has received funding from the European Union's 
Horizon 2020 research and innovation programme under grant agreement No. 
653782. We thank the OSIRIS consortium (IST/UCLA) for access to the OSIRIS code and acknowledge the computing time grant by the J\"ulich Supercomputing Center on JUQUEEN under Project No. HHH23 as well as the use of the Maxwell Cluster at DESY. We would also like to thank Dr. Reinhard Brinkmann, Dr. Klaus Floettmann and Thomas Heinemann for their relevant input and discussions.

\section*{Author contributions statement}

A.F.P. developed the analytical model, performed the simulations, prepared the figures and wrote the manuscript. A.M.O. contributed to the development of the model, provided PIC simulations support and participated in the writing of the manuscript. R.W.A. originally proposed studying the impact of the phenomena here described, provided input on the accelerator physics involved and contributed to the writing of the manuscript. 

\section*{Additional information}
\subsection*{Competing interests}
The authors declare no competing interests.

\end{document}